\documentclass [12pt,twoside] {article}
\usepackage{setspace,graphicx,natbib,bm,fancyhdr}
\usepackage{amsmath}
\usepackage{amsmath,amsthm,amssymb,latexsym}
\usepackage{color}
\usepackage{graphicx}

\usepackage{authblk}

\newcommand{\rf}{\vskip .1in\par\sloppy\hangindent=1pc\hangafter=1
                 \noindent}
\begin{document}
\title{\bf Stochastic temporal data upscaling using the generalized k-nearest neighbor algorithm}
\author{John Mashford\footnote{corresponding author, e-mail: mashford@unimelb.edu.au, tel.: +61 3 94810180, ORCID: http://orcid.org/0000-0001-6100-031X} \\
University of Melbourne \\
School of Mathematics and Statistics \\
Parkville, Vic. 3010, Australia}
\date{\today}
\maketitle
%\newcommand{\rf}{\vskip .1in\par\sloppy\hangindent=1pc\hangafter=1 \noindent}
%\newcommand{\bl}{\color{blue}}
%\raggedbottom

\begin{abstract}

Three methods of temporal data upscaling, which may collectively be called the generalized k-nearest neighbor (GkNN) method, are considered. The accuracy of the GkNN simulation of month by month yield is considered (where the term yield denotes the dependent variable). The notion of an eventually well distributed time series is introduced and on the basis of this assumption some properties of the average annual yield and its variance for a GkNN simulation are computed. The total yield over a planning period is determined and a general framework for considering the GkNN algorithm based on the notion of stochastically dependent time series is described and it is shown that for a sufficiently large training set the GkNN simulation has the same statistical properties as the training data. An example of the application of the methodology is given in the problem of simulating yield of a rainwater tank given monthly climatic data. 


\end{abstract}

\section{Introduction}

The $k$-nearest neighbor method has its origins in the work of Mack (1981), Yakowitz and Karlsson (1987) and others e.g. (Cover and Hart, 1967; Devroye, 1981). In this work an estimate for $Y_i$ given an independently and identically distributed (i.i.d.) sequence $(X_j,Y_j)$ of random vectors with $X_j\in {\bf R}^p$ and $Y_j \in {\bf R}$ (where ${\bf R}$ denotes the set of real numbers) on the basis of $\{(X_j, Y_j) : j < i\}$ is obtained by taking the average of $Y_j$ over the set $\{Y_j : j \in J\}$ where $J$ is the set of indices of vectors $X_j$ which form the $k$ nearest neighbors of $X_i$, in which $k > 1$.

In later work by Lall and Sharma (1996) and Rajagapolan and Lall (1999) a related method, also called the $k$-nearest neighbor method, was used for simulating hydrological stochastic time series $(X_i,Y_i)$. In this method the next value in the simulated time series is chosen randomly according to a probability distribution over the set $J$ of indices $j$ of the $k$ nearest neighbors $X_j$ of $X_i$ in $\{X_j : j < i\}$.

More recent work in the area has been carried out by Biau et al. (2012), Lee and Ourda (2011) and Zhang (2012).

In the present paper we derive some general results about the $k$-nearest neighbor algorithm and related methods which we group together as a general class of methods which we call the generalized $k$-nearest neighbor method (GkNN method). We do not make the assumption that the time series are i.i.d. (Mack, 1981), null-recurrent Markov (Yakowitz, 1993) or Harris recurrent Markov chains (Sancetta, 2009). We introduce the natural notion of a time series being eventually well distributed from which, if satisfied, some properties of the GkNN algorithm can be deduced.

The generalized $k$ nearest neighbor (GkNN) algorithm is described in Section 2. Section 3 investigates the problem of predicting the month by month yield (where we use the term ``yield" to denote the value of the dependent variable $Y_i$) while Section 4 considers the computation of the average annual yield. Section 5 computes the variance of the average annual yield while Section 6 considers the behavior of the total yield. Section 7 describes a general framework for viewing the GkNN algorithm and conditions under which this framework is applicable in practice. The eighth section of this paper presents the particular example of the problem of simulating rainwater tank yield. The paper concludes in Section 9.

\section{The generalized k nearest neighbor (GkNN) method} 

In the GkNN method we are given a time series $\{v_t\in V : t = 1, \ldots, T\}$ of predictor vectors  which may be obtained from, for example, a stochastic simulation of climatic data. Here $V$ denote the space of predictor vectors. We are also given training data $\{(w_i,u_i)\in V\times[0,\infty) : i = 1, \ldots, N\}$. 

We want to assign yields $y_t$ for $t = 1, \ldots, T$ in a meaningful way. We are given a metric $\mu : V\times V \rightarrow [0,\infty)$. We are also given a probability distribution $\{p_1,\ldots, p_N\}$ on $\{1, \ldots, N\}$. 
In the GkNN method the yield time series $y_t$ for t = 1, \ldots, T is computed as follows.
\newline
For each $t = 1, \ldots, T$,
\begin{enumerate}
\item Compute the metric values $\mu(v_t,w_i)$ for $i = 1, \ldots, N$ and sort from lowest to highest. Let $\pi_t$ be the resulting permutation of $\{1, \ldots, N\}$.
\item Randomly choose $i\in\{1,\ldots, N\}$ according to the distribution $\{p_1, \ldots, p_N\}$. Denote it by i\_selected.
\item Return $y_t = u_{\pi_t(\mbox{i\_selected})}$.
\end{enumerate}

\section{Prediction of the month by month yield by GkNN simulation}

We want to determine by either theoretical calculation or computational experiment how well the GkNN method predicts yields, or at least to find some sense in which it can be said that the GkNN method is predicting yields accurately. Suppose that we have a training set $\{(w_i,u_i):i=1,\ldots,N\}$. Let $\{v_t : t = 1, \ldots, T\}$ be a given climatic time series and $\{z_t : t = 1, \ldots, T\}$ associated (unknown) yields. The GkNN method is a stochastic method for generating a yield time series. Suppose that we run it $R$ times resulting in a yield time series $\{y_t^{(r)}  : t = 1, \ldots, T\}$ for run r where $r \in \{1, \ldots, R\}$. 

We will first work out how well the GkNN predicted yield approximates the actual yield for any given month. A measure of the error of the predicted yield compared to the actual yield for month $t$ and run $r$ is the square of the deviation, i.e. $(y_t^{(r)}  - z_t)^2$. The expected error for the GkNN computation of the yield for month $t$ is
\begin{equation}
E_t = \lim_{R\rightarrow\infty}\frac{1}{R}\sum_{r=1}^R (y_t^{(r)}  - z_t)^2.
\end{equation}

We will show that this expected error exists and is positive. Let $<y_t>$ denote the expected value of the GkNN prediction of the yield for month $t$. We will show that $<y_t>$ exists. Let $\gamma(t,r)$ denote the index i\_selected chosen in step 2 of the GkNN algorithm for month $t$ and run $r$. By definition
\begin{eqnarray}
<y_t> & = & \lim_{R\rightarrow\infty}\frac{1}{R}\sum_{r=1}^R y_t^{(r)} \nonumber \\
    & = & \lim_{R\rightarrow\infty}\frac{1}{R}\sum_{i=1}^N u_{\pi_t(i)}|\{r\in\{1,\ldots,R\}:\gamma(t,r)=i\}| \nonumber \\
    & = & \sum_{i=1}^N u_{\pi_t(i)}\lim_{R\rightarrow\infty}\frac{1}{R}|\{r\in\{1,\ldots,R\}:\gamma(t,r)=i\}| \nonumber \\
    & = & \sum_{i=1}^N p_i u_{\pi_t}(i). \nonumber
\end{eqnarray}
Thus $<y_t>$ exists. Now
\begin{eqnarray}
(y_t^{(r)}  - z_t)^2 & = & (y_t^{(r)}  - <y_t> + <y_t> - z_t)^2 \nonumber \\
    & = & (y_t^{(r)}-<y_t>)^2 +(<y_t>  - z_t)^2 + 2(y_t^{(r)}  - <y_t>)(<y_t>- z_t) \nonumber
\end{eqnarray}
Therefore
\begin{eqnarray}
E_t & = & \lim_{R\rightarrow\infty}\frac{1}{R}\sum_{r=1}^R (y_t^{(r)}  - <y_t>)^2 + (<y_t>-  z_t)^2 \nonumber \\
    & = & \mbox{Var}(y_t)+(<y_t>-z_t)^2 \nonumber
\end{eqnarray}
Now the variance Var$(y_t)$ is given by
\begin{eqnarray}
\mbox{Var}(y_t) & = & \lim_{R\rightarrow\infty}\frac{1}{R}\sum_{r=1}^R(y_t^{(r)}  - <y_t>)^2 \nonumber \\
    & = &  \lim_{R\rightarrow\infty}\frac{1}{R}\sum_{i=1}^N(u_{\pi_t(i)}-<y_t>)^2|\{r\in\{1,\ldots,R\}:\gamma(t,r)=i\}| \nonumber \\
    & = & \sum_{i=1}^N(u_{\pi_t(i)}-<y_t>)^2\lim_{R\rightarrow\infty}\frac{1}{R}|\{r\in\{1,\ldots,R\}:\gamma(t,r)=i\}| \nonumber \\
    & = & \sum_{i=1}^N p_i(u_{\pi_t(i)}-<y_t>)^2 \nonumber
\end{eqnarray}
Thus
\begin{equation}
E_t = \sum_{i=1}^N p_i(u_{\pi_t(i)}-<y_t>)^2+(<y_t>-z_t)^2.
\end{equation}

The expected error is the sum of two non-negative terms. The first term can only be zero if all the points in the neighborhood $\{(w_{\pi_t(i)},u_{\pi_t(i)}) : i = 1, \ldots, N; p_i > 0\}$ have associated yields equal to $<y_t>$ and this is seldom the case. The greater the distribution of yields in the neighborhood the greater the first term will be and hence, the greater $E_t$ will be. Thus the expected error $E_t$ is positive and the error in the prediction of the yield during month $t$ for any given run is likely to be positive.

A measure of the total error of the GkNN prediction of yield over the total simulation period for run $r$ is
\begin{equation}
E^{(r)} = \sum_{t=1}^T (y_t^{(r)}-z_t)^2,
\end{equation}
and its expected value is
\begin{eqnarray}
E & = & \lim_{R\rightarrow\infty}\frac{1}{R}\sum_{r=1}^R\sum_{t=1}^T(y_t^{(r)}-z_t)^2 \nonumber \\
    & = & \sum_{t=1}^T\lim_{R\rightarrow\infty}\frac{1}{R}\sum_{r=1}^R(y_t^{(r)}-z_t)^2 \nonumber \\
    & = & \sum_{t=1}^T E_t \nonumber \\
    & > & 0. \nonumber
\end{eqnarray}
We may write
\[ E_t = \sum_{i=1}^N p_i(u_{\pi_t(i)}-<y_t>)^2+(<y_t>-z_t)^2 = E_t^{(b)}+E_t^{(p)}, \]
where
\[ E_t^{(b)} =  \sum_{i=1}^N p_i(u_{\pi_t(i)}-<y_t>)^2, \]
and
\[ E_t^{(p)} = (<y_t>-z_t)^2. \]
We have
\[ E_t^{(b)} = \sum_{i=1}^Np_i(u_{\pi_t(i)}-\sum_{j=1}^N p_ju_{\pi_t(j)})^2. \]
Now define $\pi : V\times\{1,\ldots,k\}\rightarrow\{1,\ldots,N\}$ by 
\begin{eqnarray}
\pi(v,i) & = & \mbox{the index of the $i$th closest element of $\{w_j:j=1,\ldots,N\}$ to $v$} \nonumber \\
    &  & \mbox{with respect to the metric $\mu$}, \nonumber
\end{eqnarray}
and let, for $v\in V, i\in\{1,\ldots,k\}, \pi_v(i) = \pi(v,i)$.
Then 
\begin{equation}
E_t^{(b)} = E(v_t),
\end{equation}
where $E:V\rightarrow[0,\infty)$ is defined by
\begin{equation}
E(v) = \sum_{i=1}^Np_i(u_{\pi_v(i)}-\sum_{j=1}^Np_ju_{\pi_v(j)})^2.
\end{equation}
$E:V\rightarrow[0,\infty)$ may be called the base error map. We will show that E is bounded over the predictor vector space as follows.
\begin{eqnarray}
E(v) & \leq & \sum_{i=1}^Np_i|u_{\pi_v(i)}-\sum_{j=1}^Np_ju_{\pi_v(j)}|^2 \nonumber \\
    & \leq & \sum_{i=1}^Np_i(u_{\mbox{max}}+\sum_{j=1}^Np_ju_{\mbox{max}})^2 \nonumber \\
    & \leq & N(u_{\mbox{max}}+Nu_{\mbox{max}})^2 \nonumber \\
    & = & N(1+N)^2(u_{\mbox{max}})^2, \nonumber
\end{eqnarray}
 where $u_{\mbox{max}}=\mbox{max}\{u_i:i=1,\ldots,N\}$.

\section{Prediction of the annual average yield by GkNN simulation}

Thus the GkNN method does not make accurate detailed month by month predictions of the yield. We would like to determine some way in which the GkNN method gives useful information about the system behavior. We will show that under certain assumptions the GkNN method gives an accurate prediction of the average annual yield and the accuracy of the prediction increases as the total time period of the simulation increases.

Given a permutation $\pi : \{1, \ldots, N\} \rightarrow\{1, \ldots, N\}$ let $V_{\pi} = \{v\in V : \pi_v = \pi\}$. Let $\Pi$ denote the set of all permutations of $\{1, \ldots, N\}$. Suppose that the simulation is carried out over $m$ years, so $T = 12m$. The average annual yield for run $r$ is
\begin{equation}
Y^{(r)}=\frac{1}{m}\sum_{t=1}^{12m}y_t^{(r)} = \frac{1}{m}\sum_{\pi\in\Pi}\sum\{y_t^{(r)}:v_t\in V_{\pi},t\in\{1,\ldots,12m\}\}.
\end{equation}
Therefore the average of the average annual yield over $R$ runs is given by
\begin{eqnarray}
\frac{1}{R}\sum_{r=1}^R Y^{(r)} & = & \frac{1}{R}\sum_{r=1}^R\frac{1}{m}\sum_{\pi\in\Pi}\sum\{y_t^{(r)}:v_t\in V_{\pi},t\in\{1,\ldots,12m\}\} \nonumber \\
    & = & \frac{1}{m}\sum_{\pi\in\Pi}\sum\{\frac{1}{R}\sum_{r=1}^R y_t^{(r)}:v_t\in V_{\pi},t\in\{1,\ldots,12m\}\} \nonumber \\
    & \rightarrow & \frac{1}{m}\sum_{\pi\in\Pi}\sum\{\sum_{i=1}^N p_i u_{\pi(i)}:v_t\in V_{\pi},t\in\{1,\ldots,12m\}\}, \nonumber
\end{eqnarray}
as $R\rightarrow\infty$.
Therefore the expected value of the predicted average annual yield is given by
\begin{equation}
<Y>=\frac{1}{m}\sum_{\pi\in\Pi}\{\sum_{i=1}^N p_i u_{\pi(i)}|\{t\in\{1,\ldots,12m\}:v_t\in V_{\pi}\}|.
\end{equation}
If $X$ is a topological space and  and $x=\{x_t:t=1,\ldots \}$ is a time series in $X$ then we will say that $x$ is eventually well distributed if
\begin{equation}
\lim_{T\rightarrow\infty}\frac{1}{T}|\{t\in\{1,\ldots T\}:x_t\in U\}| \mbox{ exists for all Borel sets }U\subset X.
\end{equation}
(Borel$(X)=\{\mbox{Borel sets in $X$}\}$ denotes the sigma algebra generated by the set of open sets in $X$ (Halmos, 1974).) This is a natural property for a time series to have. If $x$ is eventually well distributed define its distribution to be the mapping $\nu : \mbox{Borel}(X) \rightarrow [0,1]$ defined by
\begin{equation}
\nu(U)=\lim_{T\rightarrow\infty}\frac{1}{T}|\{t\in\{1,\ldots T\}:x_t\in U\}|.
\end{equation}
It is straightforward to show that $\nu$ is finitely additive and $\nu(X) = 1$.

If the climatic time series $\{v_t : t = 1, 2, \ldots\}$ is eventually well distributed with distribution $\nu$ then the average annual yield converges to a limit as the number of years $m$ in the simulation increases given by
\begin{equation}
\lim_{m\rightarrow\infty}<Y>=12\sum_{\pi\in\Pi}\nu(V_{\pi})(\sum_{i=1}^Np_i u_{\pi(i)}).
\end{equation}

\section{Variance of the average annual yield predicted by GkNN simulation}

We will now compute the variance of the average annual yield and show that it tends to zero as the number of years $m$ in the simulation increases. We have
\begin{equation}
\mbox{Var}(Y)=\lim_{R\rightarrow\infty}\frac{1}{R}\sum_{r=1}^R(Y^{(r)}-<Y>)^2=\lim_{R\rightarrow\infty}\frac{1}{R}\sum_{r=1}^R Y^{{(r)}2}-<Y>^2. 
\end{equation}
We may compute
\begin{eqnarray}
\frac{1}{R}\sum_{r=1}^R Y^{(r)2} & = & \frac{1}{R}\sum_{r=1}^R\frac{1}{m^2}(\sum_{\pi\in\Pi}\sum\{y_t^{(r)}:v_t\in V_{\pi},t\in\{1,\ldots,12m\}\})^2 \nonumber \\ 
& = & \frac{1}{R}\sum_{r=1}^R(\frac{1}{m^2}(\sum_{\pi_1\in\Pi}\sum\{y_t^{(r)}:v_t\in V_{\pi_1},t\in\{1,\ldots,12m\}\}) \nonumber \\
    &  & (\sum_{\pi_2\in\Pi}\sum\{y_s^{(r)}:v_s\in V_{\pi_2},s\in\{1,\ldots,12m\}\})) \nonumber \\
    & =  & \frac{1}{m^2}\sum_{\pi_1,\pi_2\in\Pi}\sum\{\frac{1}{R}\sum_{r=1}^R y_t^{(r)}y_s^{(r)}:v_t\in V_{\pi_1},v_s\in V_{\pi_2},t,s\in\{1,\ldots 12m\}\} \nonumber  \\
   & =  & \frac{1}{m^2}\sum_{\pi_1,\pi_2\in\Pi}\sum\{\frac{1}{R}\sum_{r=1}^R y_t^{(r)}y_s^{(r)}:v_t\in V_{\pi_1},v_s\in V_{\pi_2},t,s\in\{1,\ldots 12m\}, \nonumber \\
    &  & s\neq t\}+\frac{1}{m^2}\sum_{\pi\in\Pi}\sum\{\frac{1}{R}\sum_{r=1}^R y_t^{(r)2}:v_t\in V_{\pi},t\in\{1,\ldots,12m\}\}. \nonumber  
\end{eqnarray}
Now for $s, t \in \{1, \ldots, 12m\}, s \neq t, v_t\in V_{\pi_1}, v_s\in V_{\pi_2}$,
\begin{eqnarray}
\frac{1}{R}\sum_{r=1}^Ry_t^{(r)}y_s^{(r)} & = & \frac{1}{R}\sum\{u_{\pi_1(i)}u_{\pi_2(j)}:\gamma(t,r)=i,\gamma(s,r)=j;i,j\in\{1,\ldots,N\}, \nonumber \\
    &  & r\in\{1,\ldots,R\}\} \nonumber \\
    & = & \frac{1}{R}\sum_{i,j=1}^N u_{\pi_1(i)}u_{\pi_2(j)}|\{r\in\{1,\ldots,R\}:\gamma(t,r)=i,\gamma(s,r)=j\}| \nonumber \\
    & = & \sum_{i,j=1}^N u_{\pi_1(i)}u_{\pi_2(j)}\frac{1}{R}|\{r\in\{1,\ldots R\}:\gamma(t,r)=i\}| \nonumber \\
    &  & |\{r\in\{1,\ldots,R\}:\gamma(t,r)=i,\gamma(s,r)=j\}|^{-1} \nonumber \\
    &  & |\{r\in\{1,\ldots,R\}:\gamma(t,r)=i,\gamma(s,r)=j\}| \nonumber \\
    & \rightarrow & \sum_{i,j=1}^Np_i p_j u_{\pi_1(i)}u_{\pi_2(j)}, \nonumber
\end{eqnarray}
as $R\rightarrow\infty$ (assuming that the index selection at step 3 of the GkNN algorithm at time $t$ is independent of its selection at time $s$). Therefore
\begin{eqnarray}
\lim_{R\rightarrow\infty}\frac{1}{R}\sum_{r=1}^R Y^{(r)2} & = & 
\frac{1}{m^2}\sum_{\pi_1,\pi_2\in\Pi}(\sum_{i,j=1}^N p_i p_j u_{\pi_1(i)}u_{\pi_2(j)})|\{(s,t)\in\{1,\ldots,12m\}^2: \nonumber \\
    &  & v_t\in V_{\pi_1},v_s\in V_{\pi_2},s\neq t\}| \nonumber +\\
    &  & \frac{1}{m^2}\sum_{\pi\in\Pi}<y_t>^2|\{t\in\{1,\ldots 12m\}:v_t\in V_{\pi}\}|. \nonumber
\end{eqnarray}
Also we compute
\begin{eqnarray}
<Y>^2   & = & \frac{1}{m^2}(\sum_{\pi\in\Pi}\sum_{i=1}^N p_i u_{\pi(i)}|\{t\in 1,\ldots 12m\}:v_t\in V_\pi\}|)^2 \nonumber \\
    & = & \frac{1}{m^2}\sum_{\pi_1,\pi_2\in\Pi}\sum_{i,j=1}^N p_i p_j u_{\pi_1(i)}u_{\pi_2(j)}|\{(s,t)\in\{1,\ldots,12m\}^2:v_t\in V_{\pi_1},v_s\in V_{\pi_2}\}|. \nonumber
\end{eqnarray}
It follows that
\begin{eqnarray}
\mbox{Var}(Y) & = & \frac{1}{m^2}\sum_{\pi\in\Pi}(<y_t^2>-<y_t>^2)|\{t\in\{1,\ldots,12m\}:v_t\in V_{\pi}\}| \nonumber \\
    & = & \frac{1}{m^2}\sum_{\pi\in\Pi}\mbox{Var}(y_t)|\{t\in\{1,\ldots 12m\}:v_t\in V_{\pi}\}| \nonumber \\
    & = & \frac{1}{m^2}\sum_{\pi\in\Pi}\sum_{i=1}^Np_i(u_{\pi(i)}-\sum_{j=1}^N p_j u_{\pi(j)})^2|\{t\in\{1,\ldots 12m\}:v_t\in V_{\pi}\}|. \nonumber
\end{eqnarray}
Therefore
\begin{equation}
\mbox{Var}(Y)\leq\frac{C}{m},
\end{equation}
where
\begin{equation}
C = 12\sum_{\pi\in\Pi}\sum_{i=1}^N p_i(u_{\pi(i)}-\sum_{j=1}^N p_j u_{\pi(j)})^2,
\end{equation} 
and so the variance of the predicted annual average annual yield as computed by the GkNN method tends to zero as the total number of years $m$ in the simulation increases. If the time series $\{v_t : t = 1, 2, \ldots \}$ is eventually well distributed with distribution $\nu$ then
\begin{equation}
\lim_{m\rightarrow\infty}(m\mbox{Var}(Y))=\sum_{\pi\in\Pi}\nu(V_{\pi})\sum_{i=1}^Np_i(u_{\pi(i)}-\sum_{j=1}^Np_j u_{\pi(j)})^2.
\end{equation}

\section{Prediction of the total yield by GkNN simulation}

Thus the computation of average annual yield using GkNN seems to be well behaved. However it is perhaps of greater interest to consider the total yield at any month starting from the beginning of the simulation period. The total yield $Y_{\mbox{tot}}$ over a simulation period of $m$ years is given by

\begin{equation}
Y_{\mbox{tot}} = mY,
\end{equation}
where $Y$ is the average annual yield. Therefore the variance of the total yield is given by
\begin{equation}
\mbox{Var}(Y_{\mbox{tot}})=m^2\mbox{Var}(Y)=\sum_{\pi\in\Pi}\sum_{i=1}^Np_i(u_{\pi(i)}-\sum_{j=1}^Np_j u_{\pi(j)})^2|\{t\in\{1,\ldots,12m\}:v_t\in V_{\pi}\}|.
\end{equation}
If the time series $\{v_t:t=1,2,\ldots\}$ is eventually well distributed with distribution $\nu$ then Var($Y_{\mbox{tot}})=mf(m)$, where
\begin{eqnarray}
f(m) &  = & \sum_{\pi\in\Pi}\sum_{i=1}^N p_i(u_{\pi(i)}-\sum_{j=1}^Np_j u_{\pi(j)})^2|\{t\in\{1,\ldots 12m\}:v_t\in V_{\pi}\}|/m \nonumber \\
    & \rightarrow & \sum_{\pi\in\Pi}\nu(V_{\pi})\sum_{i=1}^N p_i(u_{\pi(i)}-\sum_{j=1}^Np_j u_{\pi(j)})^2, \nonumber
\end{eqnarray}
as $m\rightarrow\infty$. This limit will be positive for practical applications. Thus, in this case, the variance of $Y_{\mbox{tot}}$ becomes unbounded as $m\rightarrow\infty$. 

\section{A general framework for GkNN}

Let $\{v_t : t = 1, 2, \ldots\} \subset V$ be a time series which may be a realization of some stochastic process and let $Z$ be a topological space. A stochastic process $y=\{y_t : t = 1, 2, \ldots\}\subset Z$ will be said to be stochastically dependent on $\{v_t:t=1,2\ldots\}$ if there exists a continuous kernel $K : V \times \mbox{Borel}(Z) \rightarrow [0,1]$ such that
\begin{equation} \label{eq:gen1}
\mbox{Pr}(y_t \in\Gamma) = K(v_t,\Gamma), \forall t = 1, 2, \ldots
\end{equation}                                                                            
The condition that $K$ is a continuous kernel means that for all $v \in V$ the mapping taking $\Gamma\in\mbox{Borel}(Z)$ to $ K(v,\Gamma)$ is a probability measure and for all $\Gamma\in \mbox{Borel}(Z)$ the mapping taking $v \in V$ to $K(v,\Gamma)$ is continuous. Equation \ref{eq:gen1} means that if $\{y_t^{(r)}  : t = 1, 2, \ldots\}$ for $r = 1, 2, \ldots$ are a collection of runs (replicates) of the stochastic process $\{y_t : t = 1, 2, \ldots\}$ then
\begin{equation}
\lim_{R\rightarrow\infty}\frac{1}{R}|\{r \in \{1, \ldots, R\} : y_t^{(r)}\in\Gamma\}| =  K(v_t,\Gamma).
\end{equation}
Consider the GkNN process defined by training data $\{(w_i,u_i) : i = 1, \ldots, N\} \subset V \times[0,\infty)$. In this case the space $Z$ is the space $[0,\infty)$. We will show tha the process $\{y_1,y_2,\ldots\}$ is stochastically dependent on the time series $\{v_t:t=1,2,\ldots\}$. In fact we have that
\begin{eqnarray}
\mbox{Pr}(y_t\in\Gamma) & = & \lim_{R\rightarrow\infty}\frac{1}{R}|\{r\in\{1,\ldots,R\}:y_t^{(r)}\in\Gamma\}| \nonumber \\
    & = & \lim_{R\rightarrow\infty}\frac{1}{R}\sum_{i=1}^N\delta_{u_{\pi_t(i)}}(\Gamma)|\{r\in\{1,\ldots,R\}:y_t^{(r)}=u_{\pi_t(i)}\}| \nonumber \\
    & = & \sum_{i=1}^N\delta_{u_{\pi_t(i)}}(\Gamma)\lim_{R\rightarrow\infty}\frac{1}{R}|\{r\in\{1,\ldots,R\}:y_t^{(r)}=u_{\pi_t(i)}\}| \nonumber \\
    & = & \sum_{i=1}^N p_i \delta_{u_{\pi_t(i)}}(\Gamma), \nonumber
\end{eqnarray}
where, for $a\in Z$, $\delta_a:\mbox{Borel}(Z)\rightarrow[0,\infty)$ denotes the Dirac measure concentrated on $a$ defined by
\begin{equation}
\delta_a(\Gamma)=\left\{\begin{array}{l}
1 \mbox{ if } a\in\Gamma \nonumber \\
0 \mbox{ otherwise}
\end{array}\right.
\end{equation}
It follows that the GkNN process is stochastically dependent on $\{v_t:t=1,2,\ldots\}$ with kernel $K$ defined by
\begin{equation}
K(v,\Gamma)=\sum_{i=1}^N p_i\delta_{u_{\pi_v(i)}}(\Gamma).
\end{equation}
Now suppose that $\{v_t : t = 1, 2, \ldots\} \subset V$ is a time series and $\{y_t : t = 1, 2, \ldots\}\subset [0,\infty)$ is a stochastic process which is stochastically dependent on $\{v_t : t = 1, 2, \ldots\}$ with kernel $K$ where $K$ is defined by a continuous functional kernel $\phi : V \times [0,\infty) \rightarrow [0,\infty)$, i.e.
\begin{equation}
K(v,\Gamma)=\int_{\Gamma}\phi(v,\xi)\,d\xi.
\end{equation}
Let $\{z_t : t = 1, 2, \ldots\}$ be a realization (replicate) of $\{y_t : t = 1, 2, \ldots\}$ and let $K_N$ be the kernel associated with the GkNN process with training set $W_N = \{(v_i,z_i) : i = 1, \ldots, N\}$ and probabilities
\begin{equation}
p_i=\left\{\begin{array}{l}
\frac{1}{k_N}\mbox{ for }i=1,\ldots,k_N \nonumber\\
0\mbox{ otherwise}
\end{array}\right.
\end{equation}
for which $k_N\rightarrow\infty$ as $N\rightarrow\infty$ but $\frac{k_N}{N}\rightarrow 0$ as $N\rightarrow\infty$. An example of a sequence $k_N$ satisfying this is $k_N=\sqrt{N}$. $K_N$ is given by
\begin{equation}
K_N(v,\Gamma)=\frac{1}{k_N}\sum_{i=1}^{k_N}\delta_{z_{\pi_v(i)}}(\Gamma)=\frac{1}{k_N}|\{i\in\{1,\ldots,k_N\}:z_{\pi_v(i)}\in\Gamma)\}|.
\end{equation}
Therefore for an interval $(a,b)$
\begin{equation}
K_N(v,(a,b))=\frac{1}{k_N}|\{i\in\{1,\ldots,k_N\}:z_{\pi_v(i)}\in(a,b)\}|.
\end{equation} 
Now let $\psi:V\times[0,\infty]\rightarrow[0,1]$ be defined by
\begin{equation}
\psi(v,\xi)=\int_0^{\xi}\phi(v,\zeta)\,d\zeta.
\end{equation}
Let $\{\rho_t\}$ be defined by $\rho_t=\psi(v_t,z_t)$ for $t=1,2,\ldots$. Then $\{\rho_t\}$ is a uniformly distributed sequence of random numbers and $z_t=(\psi(v_t,\mbox{ . }))^{-1}(\rho_t)$. Thus
\begin{eqnarray}
K_N(v,(a,b)) & = & \frac{1}{k_N}|\{i\in\{1,\ldots,k_N\}:(\psi(v_{\pi_v(i)},\mbox{ . }))^{-1}(\rho_{\pi_v(i)})\in(a,b)\}| \nonumber \\
    & = & \frac{1}{k_N}|\{i\in\{1,\ldots,k_N\}:(\psi(v,\mbox{ . }))^{-1}(\rho_{\pi_v(i)})\in(a,b)\}| \nonumber \\
    & = & \frac{1}{k_N}|\{i\in\{1,\ldots,k_N\}:\rho_{\pi_v(i)}\in(\psi(v,a),\psi(v,b))\}| \nonumber \\
    & \rightarrow & \psi(v,b)-\psi(v,a) \nonumber \\
    & =&  K(v,(a,b)), \nonumber
\end{eqnarray}
as $N \rightarrow\infty$, assuming that $\mu(v_{\pi_v(i)},v)$ is small for all $i = 1, \ldots, k_N$ for $N$ large enough (this will follow, if $\{v_t\}$ is eventually well distributed with positive distribution, given that $\{\rho_{\pi_v(i)}\}$ is a uniformly distributed sequence).

Thus the GkNN kernel equals the kernel of the dependent process in the sense defined above as long as the training set for the GkNN process is large enough.

\section{Example of temporal upscaling of (rainwater) tank data}

We would like to estimate the month by month yield of a rainwater tank (RWT) given monthly climatic data. This is not straightforward because a monthly time step is too coarse for the RWT simulation model. To obtain reasonably accurate results a daily time step must be used for the RWT simulation (Mashford and Maheepala, 2015; Mashford {\em et al.}, 2011).

The monthly climatic data arises from the water supply headworks (WSH) model (Cui and Kuczera, 2003) and is usually stochastically generated with a very large time span (e.g. 1,000,000 years). The problem of temporal scaling up would not arise if the climatic data for the WSH model had a daily time step (and also if the RWT simulation algorithm could be executed sufficiently fast). 

Temporal downscaling has been used extensively in studying the short term effects of long term climate models such as models of climate change (Gangopadhyay {\em et al.}, 2005; Fowler {\em et al.}, 2007; Maraun {\em et al.}, 2010; Erhardt {\em et al.}, 2015). However in the present paper we are considering the problem of upscaling relatively short records of daily data to generate long term records of monthly data.

Three methods of temporal upscaling are the nearest neighbor (NN) method of Coombes et al. (2002), Kuczera's bootstrap method (Kuczera, 2008) and the $k$ - nearest neighbor (kNN) method (Lall and Sharma, 1996; Rajagapolan and Lall, 1999; Gangopadhyay {\em et al.}, 2005). 

In each of these methods the RWT month by month yield associated with a WSH climatic time series is estimated using a comparatively short (e.g. 140 years) historical record of daily climatic data. In each case the RWT simulation model (or, more generally, the Allotment Water Balance model described in (Coombes {\em et al}, 2002) is run on this daily historical record for various RWT parameter settings. In order to do this it is necessary to have a demand model which is either a simulation or, as is unlikely, a historical record. The demand simulation will take into account the climatic variables, in particular, the temperature.

The upscaling methods can be described in terms of the following general format. Each of the upscaling methods aggregates the daily RWT yields and climatic variables obtained from running the RWT simulation on the historical record into monthly time steps. They then generate a list $\{r_j^R  : j = 1, \ldots, N\}$ of records of the form
\begin{eqnarray}
r_j^R & = & (\mbox{month\underline{ }label}_ j^R, \mbox{climatic\_variable\_1}_j^R, \ldots, \mbox{climatic\_variable\_n}_j^R,\nonumber \\
    &  & \mbox{RWT\_yield}_j^R), 
\end{eqnarray}
where $N$ is the number of months in the historical record. The month label is a number in $\{1, \ldots, 12\}$ determined from the month corresponding to the record. For the method described in (Coombes {\em et al.}, 2002), $n = 3$ and 
\newline
climatic\_variable\_1 = average\_temperature,
\newline
climatic\_variable\_2 = number\_of\_rainfall\_days,
\newline
climatic\_variable\_3 = rainfall\_depth.

For Kuczera's bootstrap method and the kNN method as currently implemented $n = 1$ and
climatic\_variable\_1 = rainfall\_depth.

Now for all three upscaling methods we are given a sequence $c_1^H, c_2^H , \ldots $ of monthly records coming from the WSH model where

$c_i^H$  = (month\_label$_i^H$ , climatic\_variable\_1$_i^H$ , \ldots, climatic\_variable\_n$_i^H$ ).

For each $i$ we want to select a RWT yield to associate with $c_i^H$ . The NN method does this by finding the record in $\{r_j^R  : j = 1, \dots, N, \mbox{month\_label}_j^R = \mbox{month\_label}_i^H \}$ which is closest to $c_i^H$  as measured by the metric (a variant of the Manhattan metric) given by
\begin{equation}
\mu_{NN}(r_j^R,c_i^C)=\sum_{p=2}^d w_p|(r_j^R)_p-(c_i^H)_p|,
\end{equation}
where $d=n+1$ is the record length (e.g. 2) and $w_1, \ldots, w_n$ are weights which were chosen to be 1 in (Coombes {\em et al.}, 2002). The NN method is deterministic.

The kNN method is a stochastic method in which the following steps are carried out. 
\begin{enumerate}
\item Evaluate the distance from each record $r_j^R$  to $c_i^H$  using the following metric (a variant of the Euclidean metric) \newline
\begin{equation}
\mu_{NN}(r_j^R,c_i^H)=(\sum_{p=1}^d [((r_j^R)_p-(c_i^H)_p)/s_p]^2)^{\frac{1}{2}},
\end{equation}
where $s_p$ is the standard deviation of $\{(r_j^R)_p:j=1,\ldots, N\}$.
\item Sort the metric values
\item Choose the top (closest) $k$ values $r_{j_1}^R$ , \ldots, $r_{j_k}^R$ 
\item Assign a probability to each of the $k$ selected values proportional to $1/t$ for $t = 1, \ldots, k$
\item Randomly select an index $t$ according to the assigned probabilities and return the RWT\_yield$^R_{j_t}$ as the RWT yield corresponding to $c_i^H$ 
\end{enumerate} 
The bootstrap method is a stochastic method in which a scatter plot of \newline $\{(\mbox{rainfall}_j^R , \mbox{RWT\_yield}_j^R) : j = 1, \ldots, N\}$ is created. The domain of the plot is divided up into bands of 50 samples per band. Then, given a WSH climatic record $c_i^H$  the corresponding RWT yield is obtained by finding the band containing $\mbox{rainfall}_i^H$, randomly choosing a sample in that band and then returning its RWT yield value.

The bootstrap method of Kuczera can be modified by taking the band of samples associated with any given rainfall value to be the set of samples whose rainfall values are the 50 closest values to the given rainfall value rather than using predefined bands of 50 rainfall values. It can be argued that the modified bootstrap method is superior to the bootstrap method because the closest values are the most appropriate values to use and, for example, if the given rainfall value falls near the boundary of one of the predefined bands then the predicted yield using the bootstrap method will be biased towards the values near the centre of the band.

The modified bootstrap method, the Coombes method and the kNN method are all examples of the GkNN method.  For the modified bootstrap method the predictor vectors have one component, the rainfall. For the Coombes method the predictor vectors have three components, the average temperature, the number of rainfall days and the rainfall depth. For the kNN method the predictor vectors have two components, the month label (an integer in $\{1, \ldots, 12\}$) and the rainfall depth. The training data is obtained by running the RWT simulation model using a daily time step over a relatively short period of time (e.g. 100 years) and then upscaling to a monthly time step by aggregation. The GkNN  metric $\mu : V\times V \rightarrow [0,\infty)$ may be the modified Manhattan metric of the Coombes method or the modified Euclidean metric of the kNN.

For the bootstrap method the probability distribution on the set of nearest neighbors is given by
\begin{equation}
p_i = \left\{\begin{array}{l}
\frac{1}{50} \mbox{ for }i=1,\ldots 50 \\
0 \mbox{ otherwise} 
\end{array}\right.
\end{equation}
For the kNN method the distribution is given by
\begin{equation}
p_i = \nu^{-1}\left\{\begin{array}{l}
\frac{1}{i} \mbox{ for }i=1,\ldots k \\
0 \mbox{ otherwise}  
\end{array}\right.
\end{equation}
where
\[ \nu = \sum_{i=1}^k\frac{1}{i}. \]

\section{Conclusion}

A generalization of three methods of temporal data upscaling, which we have called the generalized k-nearest neighbor (GkNN) method, has been considered. The accuracy of the GkNN simulation of month by month yield has been considered. The notion of an eventually well distributed time series is introduced and on the basis of this assumption some properties of the average annual yield and its variance for a GkNN simulation are computed. The behavior of the total yield over a planning period has been described. A general framework for considering the GkNN algorithm based on the notion of stochastically dependent time series has been described and it is shown that for a sufficiently large training set the GkNN simulation has the same statistical properties as the training data. An example of the application of the methodology has been given in the problem of simulating the yield of a rainwater tank given monthly climatic data. 
 
\section*{Acknowledgements}

The work described in this paper was partially funded by the Commonwealth Scientific and Industrial Research Organisation (CSIRO, Australia). Also the author would like to thank Fareed Mirza, Shiroma Maheepala and Yong Song for very helpful discussions. The author is also very grateful to the anonymous reviewer for his/her very helpful suggestions.

\section*{References}

\rf  Biau, G., Devroye, L., Dujmovic, V. and Krzyak, A., ``An affine invariant k-nearest neighbour regression estimate", Journal of Multivariate Analysis 112, 2012, 24-34.

\rf Coombes, P. J., Kuczera, G., Kalma, J. D. and Argue, J. R., ``An evaluation of the benefits of source control measures at the regional scale", Urban Water 4, 2002, 307-320.

\rf Cover, T. M. and Hart, P. E., ``Nearest neighbour pattern classification", IEEE Transactions on Information Theory 13(1), 1967, p. 21.

\rf Cui, L. J., Kuczera, G, ``Optimizing urban water supply headworks using probabilistic search methods", Journal of Water Resources Planning and Management - ASCE 129(5), 2003, 380-387.

\rf Devroye, L. P., ``On the almost everywhere convergence of nonparametric regression function estimates", Annals of Statistics 9, 1981, 1310-1319.

\rf Erhardt, R. J., Band,  L. E., Smith,  R. L., Lopes, B. J., ``Statistical downscaling of precipitation on a spatially dependent
network using a regional climate model", Stoch Environ Res Risk Assess, 2015, 29, 1835-1849. 

\rf Fowler H., Blenkinsop, S., Tebaldi, C., ``Linking climate change
modelling to impact studies: recent advances in downscaling
techniques for hydrological modelling". Int J Climatol, 2007,
27, 1547-1578.

\rf Gangopadhyay, S., Clark, M., Rajagopalan, B,.``Statistical
downscaling using K-nearest neighbors". Water Resour. Res., 2005,
41, W02024.

\rf Halmos, P. R., ``Measure Theory", Springer, New York, 1974.

\rf Kuczera, G., ``Urban water supply drought security: A comparative analysis of complimentary centralised and decentralised storage systems", Proc. Water Down Under 2008, 2008, 1532-1543.

\rf Lall, U. and Sharma, A., ``A nearest neighbour bootstrap for resampling hydrologic time series", Water Resources Research 32(3), 1996, 679-693.

\rf Lee, T. and Ouarda, T. B. M. J., ``Identification of model order and number of neighbors for k-nearest neighbour resampling", Journal of Hydrology 404, 2011, 136-145.

\rf Mack, Y. P., ``Local properties of k-NN regression estimates", SIAM Journal of Algebraic and Discrete Methods 2(3), 1981.

\rf Maraun, D., Wetterhal, I. F., Ireson, A., Chandler, R., Kendon E., Widmann,
M., Brienen, S., Rust, H., Sauter, T., Themesl, M., Venema, V., Chun,  K., Goodess, C., Jones, R., Onof, C., Vracv, M., Thiele-Eich, I., 
``Precipitation downscaling under climate change: recent developments
to bridge the gap between dynamical models and the
end user". Rev Geophys {\bf 2010} 48, RG3003.

\rf Mashford, J. and Maheepala, S., ``A general model for the exact computation of yield from a rainwater tank", Applied Mathematical Modelling 39(7), 2015, 1929-1940.

\rf Mashford, J., Maheepala, S. Neumann, L. and Coultas, E., ``Computation of the expected value and variance of the average annual yield for a stochastic simulation of rainwater tank clusters", Proc. of the 2011 International Conference on Modeling, Simulation and Visualization Methods, Las Vegas, USA, 2011, 303-309.

\rf Rajagopalan, B. and Lall, U., ``A k-nearest neighbour simulator for daily precipitation and other whether variables", Water Resources Research 35(10), 1999, 3089-3101.

\rf Sancetta, A.,``Nearest neighbour conditional estimation for Harris recurrent Markov chains", Journal of Multivariate Analysis 100(10), 2009, 2224-2236.

\rf Yakowitz, S., ``Nearest neighbour regression estimation for null-recurrent Markov time series", Stochastic Processes and their Applications 48, 1993, 311-318.

\rf Yakowitz, S. and Karlsson, M., ``Nearest neighbour methods for time series, with application to rainfall/runoff prediction", in Stochastic Hydrology, Macneill, J. B. and Umphrey, G. J. (eds.), D. Reidel Publishing Company, 1987, 149-160.  

\rf Zhang, S., ``Nearest neighbour selection for iteratively kNN imputation", Journal of Systems and Software 85, 2012, 2541-2552.

\end{document}